\documentclass[prb,twocolumn,showpacs,preprintnumbers,amssymb,superscriptaddress]{revtex4}

\usepackage{graphicx}
\usepackage[tbtags]{amsmath}
\usepackage{dcolumn}
\usepackage{bm}

\begin{document}

\title
{Voltage-current and voltage-flux characteristics of asymmetric
high $T_C$  \textbf{DC SQUIDs}.}

\author{I. L. Novikov}
\affiliation{%
Novosibirsk State Technical University, 20 K. Marx
Ave., 630092 Novosibirsk, Russia.}%
\author{Ya. S. Greenberg}
\affiliation{%
Novosibirsk State Technical University, 20 K. Marx
Ave., 630092 Novosibirsk, Russia.}%
\author{V. Schultze}
\affiliation{%
Institute of Photonic Technology, Jena,
Germany}%
\author{R. IJsselsteijn}
\affiliation{%
Institute of Photonic Technology, Jena,
Germany}%
\author{H.-G. Meyer}
\affiliation{%
Institute of Photonic Technology, Jena,
Germany}%

\date{\today}
\begin{abstract}

We report measurements of transfer functions and flux shifts of 20
on-chip high T$_C$ DC SQUIDs half of which were made purposely
geometrically asymmetric. All of these SQUIDs were fabricated
using standard high T$_C$ thin film technology and they were
single layer ones, having 140 nm thickness of
YBa$_2$Cu$_3$O$_{7-x}$ film deposited by laser ablation onto MgO
bicrystal substrates with 24$^0$ misorientation angle. For every
SQUID the parameters of its intrinsic asymmetry, i. e., the
density of critical current and resistivity of every junction,
were measured directly and independently. We showed that the main
reason for the on-chip spreading of SQUIDs' voltage-current and
voltage-flux characteristics was the intrinsic asymmetry. We found
that for SQUIDs with a relative large inductance ($L>120 $ pH)
both the voltage modulation and the transfer function were not
very sensitive to the junctions asymmetry, whereas SQUIDs with
smaller inductance ($L\simeq 65-75 $ pH) were more sensitive. The
results obtained in the paper are important for the implementation
in the sensitive instruments based on high T$_C$ SQUID arrays and
gratings.

\end{abstract}


\maketitle

\section{Introduction}
The high-transition-temperature superconducting quantum
interference devices (high-T$_C$ DC SQUIDs) each consisting of two
bicrystal Josephson junctions are key elements for many sensitive
instruments such as extremely low noise magnetometers\cite{Koelle,
Lud, Fal}, the series- and parallel- SQUID arrays\cite{Schul,
Mat}, the superconducting quantum interference grating\cite{Mil,
Schul1}, etc.

However, the further developments of these devices are limited by
significant on chip spreading in the critical current and normal
resistance of high $T_C$ Josephson junctions which seems to be
unavoidable for grain-boundary junctions.\cite{Gerd, Shadr1,
Shadr2, Jeng1, Jeng2}.

The different values of critical current and normal resistance of
two Josephson junctions result in turn in the vast spreading of
the output voltage-current (VCC) and voltage flux (VFC)
characteristics for on chip high T$_C$ DC SQUIDs\cite{Beyer, Park,
Jia}.

The influence of the junction asymmetry on the output
characteristics of high $T_C$ DC SQUIDs has been analyzed
in\cite{Kleiner, Mueller, Testa1, Testa2, Green1}, where it was
shown that the transfer function for asymmetric SQUIDs can be
substantially differ from that of symmetric SQUIDs.

The extensive comparison of experimental characteristics of
intentionally fabricated asymmetric high $T_C$ DC SQUID with
computer simulations has been performed in \cite{Mueller}, where
in most cases a reduction of the voltage-to-flux transfer function
of the asymmetric SQUID as compared to the symmetric SQUID has
been observed.

From the other point the output characteristics of high $T_C$ DC
SQUIDs can be used to infer the information about the on chip
distribution of the critical currents and normal resistances of
the Josephson junctions\cite{Jeng1, Jeng2}.

In known experiments\cite{Mueller, Jeng1, Jeng2} the asymmetry of
critical current of two Josephson junctions in DC SQUID loop has
been determined indirectly from the measured flux shift of the
voltage-flux curve. Since for high $T_C$ Josephson junctions the
density of critical current and the resistivity of the junction
are interrelated, it also influences the value of the last
quantity. In addition, as is shown in the paper, this method has a
restricted range of validity. That is why, as was noted
in\cite{Mueller}, a clear-cut comparison between simulation and
experimental data requires the independent experimental
determination of the current and resistance asymmetry.

Therefore, the main purpose of our paper is the independent and
direct determination of the current and resistance asymmetry and
the investigation of the influence of the junctions asymmetry on
the scattering over the chip of the main output parameters of DC
SQUID, i.e., its voltage modulation, voltage-to-flux transfer
function, flux shift.

To this end we investigated 20 purposely made asymmetric high
$T_C$ DC SQUIDs. For the first time we measured independently and
directly the current and resistance asymmetry for every junction
in the interferometer loop. We show that the main reason for the
junctions asymmetry is the on-chip spreading in the resistivity
and the critical current density. Even for a geometrically
symmetric design of the interferometer the critical current
densities of the two junctions may be substantially different. We
found that, in general, both the voltage modulation and the
transfer function are not very sensitive to the junctions
asymmetry, whereas the flux shift of the VFC shows approximately
linear dependance on asymmetry parameters.

The paper is organized as follows. In Section II the current and
resistance asymmetry for the junctions of DC SQUID are defined and
the analytical expressions for transfer function and flux shift
for asymmetric DC SQUID are given. The experimental part of the
paper is in detail described in Section III. Section IV is devoted
to an extensive analysis of the influence of SQUID asymmetry on
the transfer function and flux shift of VFC. Obtained results are
shortly summarized in Section V.

\section{Asymmetric DC SQUID}
\subsection{The asymmetry parameters}
\begin{figure}
  \includegraphics[width=6 cm, angle=-90]{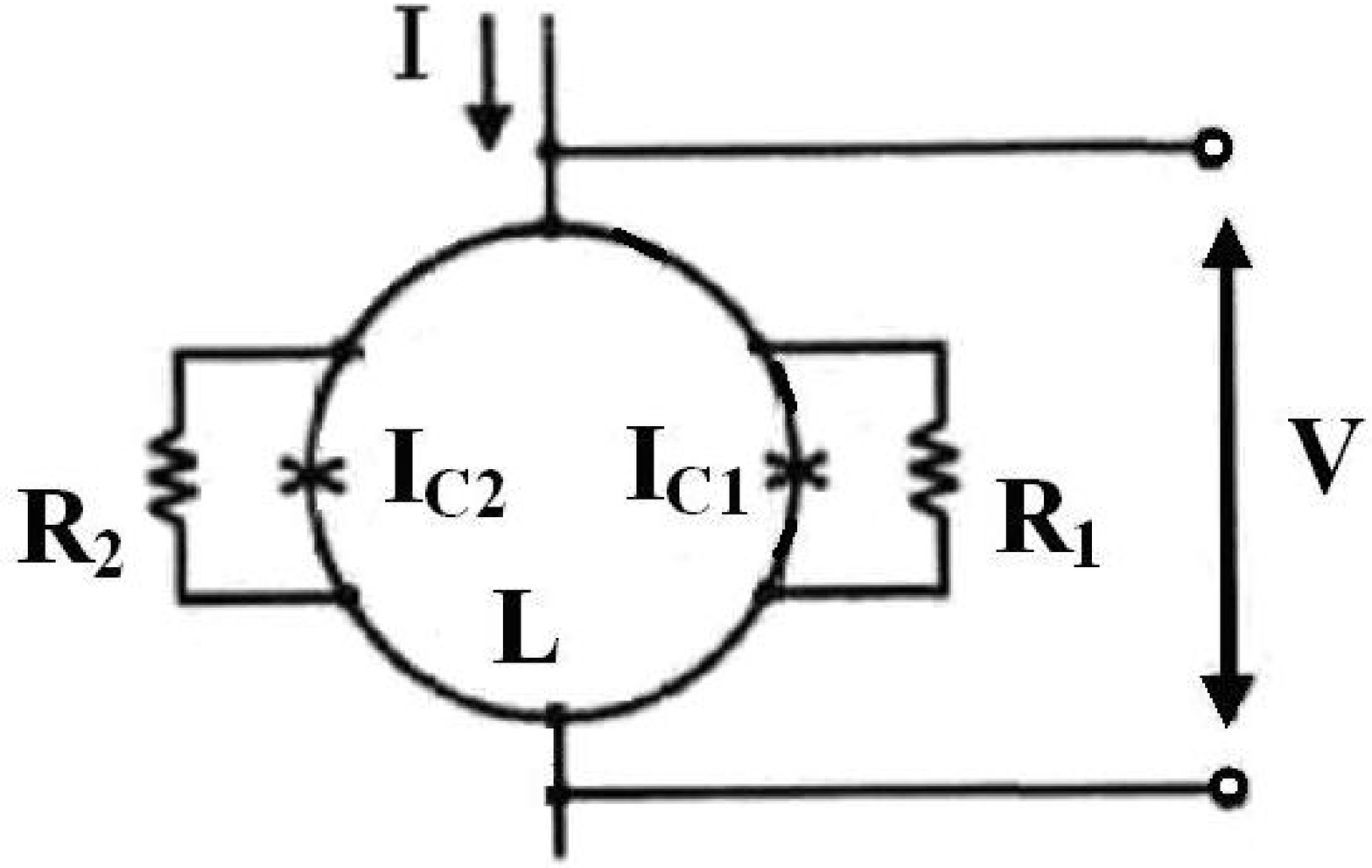}\\
  \caption{The asymmetric DC SQUID}\label{Fig1}
\end{figure}

An asymmetric DC SQUID shown in Fig.\ref{Fig1} consists of a
superconducting loop of inductance $L$ intersected by two
Josephson junctions, which have different critical currents
$I_{C1}$, $I_{C2}$ and normal resistances $R_1$, $R_2$. The
capacitances of the two junctions, which are not shown in
Fig.\ref{Fig1} are assumed to be equal.
In order to describe the junctions asymmetry we define the average
values of the critical current $I_C$ and resistance $R$, and a
current asymmetry $\gamma$ and a resistance asymmetry $\rho$ as
follows: $I_{C1}=(1+\gamma)I_C$, $I_{C2}=(1-\gamma)I_C$,
$R_1=R/(1+\rho)$, $R_2=R/(1-\rho)$, where
\begin{equation}\label{curr}
    I_C=\frac{I_{C1}+I_{C2}}{2}\equiv\frac{I_{SQ}}{2}, \texttt{      }  \gamma=\frac{I_{C1}-I_{C2}}{I_{C1}+I_{C2}}
\end{equation}
\begin{equation}\label{resist}
   R=\frac{2R_1R_2}{R_1+R_2}\equiv 2R_{SQ}, \texttt{      }  \rho=\frac{R_2-R_1}{R_1+R_2}
\end{equation}
Note that the SQUID critical current $I_{SQ}$ and its resistance
$R_{SQ}$ are being measured directly from the voltage-current
characteristic of the DC SQUID.

There are two different origins of the junction asymmetry:
geometrical asymmetry and intrinsic asymmetry\cite{Mueller}. For
bicrystal grain boundary junctions geometric asymmetry is
associated with the different width $w$ of the junctions. We
describe geometric asymmetry by the parameter $\alpha_g$ according
to $w_1=(1+\alpha_g)w$ and $w_2=(1-\alpha_g)w$, with
$w=(w_1+w_2)/2$. The intrinsic asymmetry is associated with
different values of the current density $j_0$ and the resistivity
$\rho_0$ for two junctions. We describe intrinsic asymmetry by the
parameters $\alpha_j$ and $\alpha_{\rho}$ according
to\cite{Mueller}:

\begin{multline}
j_{01}=j_0(1+\alpha_j),\texttt{          }\rho_1=\rho_0/(1+\alpha_{\rho})\\
j_{02}=j_0(1-\alpha_j),\texttt{  } \rho_2=\rho_0/(1-\alpha_{\rho})
\end{multline}
The parameters of intrinsic asymmetry $\alpha_j$ and
$\alpha_{\rho}$ can be obtained from the independent measurements
of the parameters of the bulk asymmetry $\gamma$, $\rho$ and
geometric asymmetry $\alpha_g$\cite{Mueller}:
\begin{equation}\label{intr asymm}
    \alpha_j=\frac{\gamma-\alpha_g}{1-\alpha_g\gamma},\texttt{       }
     \alpha_{\rho}=\frac{\rho-\alpha_g}{1-\alpha_g\rho}
\end{equation}
Since high $T_C$ Josephson junctions obey the scaling law
$I_CR\approx j_c^{1/2}$\cite{Gross}, the parameters $\alpha_j$ and
$\alpha_{\rho}$ are interrelated to each other\cite{Mueller}
\begin{equation}\label{intr asymm rel}
    \alpha_{\rho}=\frac{1-\sqrt{1-\alpha_j^2}}{\alpha_j},\texttt{     }
     \alpha_j=\frac{2\alpha_{\rho}}{1+\alpha_{\rho}^2}
\end{equation}
\subsection{Output voltage across asymmetric SQUID}
In general, the output voltage across asymmetric SQUID is the
complicated function of several dimensionless
parameters\cite{Green1}: $V=F(i, \alpha, \beta, \Gamma, \varphi_X,
\rho, \gamma)$, where $i=I/I_C$ ($I$ is the bias current);
$\alpha=L/L_F$ ($L_F=\left(\Phi_0/2\pi\right)^2/k_BT$ is the
fluctuation inductance which is equal approximately to 100 pH at
T=77 K); $\beta=2 LI_C/\Phi_0$; $\Gamma=2\pi k_BT/\Phi_0I_C$ is
the noise parameter ($k_B$ is Boltzmann constant, $T$ is absolute
temperature); $\varphi_X=\pi\Phi_X/\Phi_0$ ($\Phi_X$ is the
external magnetic flux, $\Phi_0$ is the flux quantum). Three
parameters $\alpha, \beta, \Gamma$ are not independent, but are
subject to the relation $\alpha=\pi\beta\Gamma$.

There are several properties of the output characteristics of
asymmetric SQUIDs which allow one to identify them by
experiment\cite{Green1}. First, in the presence of applied flux
the voltage across asymmetric SQUIDs is not an odd function of the
bias current, i. e., $V(-i)\neq-V(i)$. Thus, the quantity
$V(-i)+V(i)$ can be used as one of the measure of SQUID asymmetry.
It is a periodic function of the applied flux and it depends on
the bias current and on the parameters of asymmetry.


The second property which is known for a long time (see, for
example, Ref.\onlinecite{Ramos}) is the shift of the voltage-flux
characteristic (VFC) under reversal of the bias current. The shift
is commonly attributed to the asymmetry in the critical current of
the two junctions:
\begin{equation}\label{shift1}
    \Delta\Phi=L(I_{C1}-I_{C2})=\gamma\beta\Phi_0
\end{equation}
This property has been widely used for experimental determination
of the current asymmetry \cite{Mueller, Weiss, Jeng1, Jeng2}.
However, the expression (\ref{shift1}) cannot describe some
experimental facts. In particular, it does not depend on the bias
current while the experimental flux shift does depend on $i$. For
large inductance DC SQUIDs the picture is more complicated
\cite{Green1}. There is no simple relation between current
asymmetry and a flux shift. In addition, asymmetry in resistance
contributes also to the total flux shift. The expression for the
flux shift under a reversal of the bias current is as
follows\cite{Green1}:

\begin{equation}\label{shift3}
    \frac{{\Delta \Phi _X }}{{\Phi _0 }} = \frac{1}{{\pi }}\Theta(i, \Gamma, \rho, \gamma)
     +\frac{{\rho \alpha }}{{2\pi \Gamma }}i
\end{equation}

The explicit expression for the quantity $\Theta$ which is too
cumbersome one can find in Ref. \onlinecite{Green1} (Eq. 31b).

As is seen from (\ref{shift3}) the flux shift is a complicated
function of the bias current and the parameters of asymmetry. The
dependance on SQUID inductance $L$ is hidden in the second term in
(\ref{shift3}). Due to the relation $\alpha=\pi\beta\Gamma$ this
term can be written as $\rho i\beta/2$, which is equivalent to the
expression found for the flux shift in the limit of a large bias
current $i$ \cite{Clarke}.

It is worth noting that for $i>0$ the voltage-flux curve is
shifted to the negative side of the flux ($\Delta\Phi_X<0$)
relative to a symmetric voltage-flux curve, while for the reversed
bias ($i<0$) the voltage-flux curve is shifted to the positive
side of the flux ($\Delta\Phi_X>0$).

\subsection{Transfer function of asymmetric SQUID}

As is known, the shape of experimental VFC is not symmetric. The
asymmetry of the junctions induces distortion of the
$V(\Phi)$-curves, which leads to different values of $V_{\Phi}^+$
and $V_{\Phi}^-$ for the maximum positive and negative slope of
the $V(\Phi)$-curves, respectively. The impact of the asymmetry on
the transfer function has been studied by computer simulations in
Ref. \onlinecite{Mueller}. It was shown that the normalized
transfer function $v^{\pm}_{\Phi}=V^{\pm}_{\Phi}\Phi_0/RI_C$ can
be factorized as $v^{\pm}_{\Phi}=g^{\pm}(\Gamma\beta)
f^{\pm}(\gamma,\beta)$. The explicit analytical expressions for
the quantities $g^{+}(\Gamma\beta)$  and $f^{+}(\gamma,\beta)$ for
limited range of parameters $\Gamma$ and $\beta$ ($\Gamma\beta<1$,
$\beta<5$) have been given in Ref. \onlinecite{Mueller}.

\section{Experimental}
\subsubsection{High-T$_C$ DC SQUID preparation and characterization}

\begin{figure}
  \includegraphics[width=8 cm]{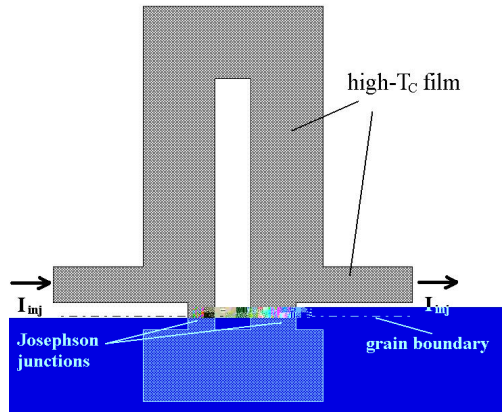}\\
  \caption{Design of asymmetric high-T$_C$ DC SQUID}\label{design}
\end{figure}

\begin{figure}
  \includegraphics[width=7 cm, angle=-90]{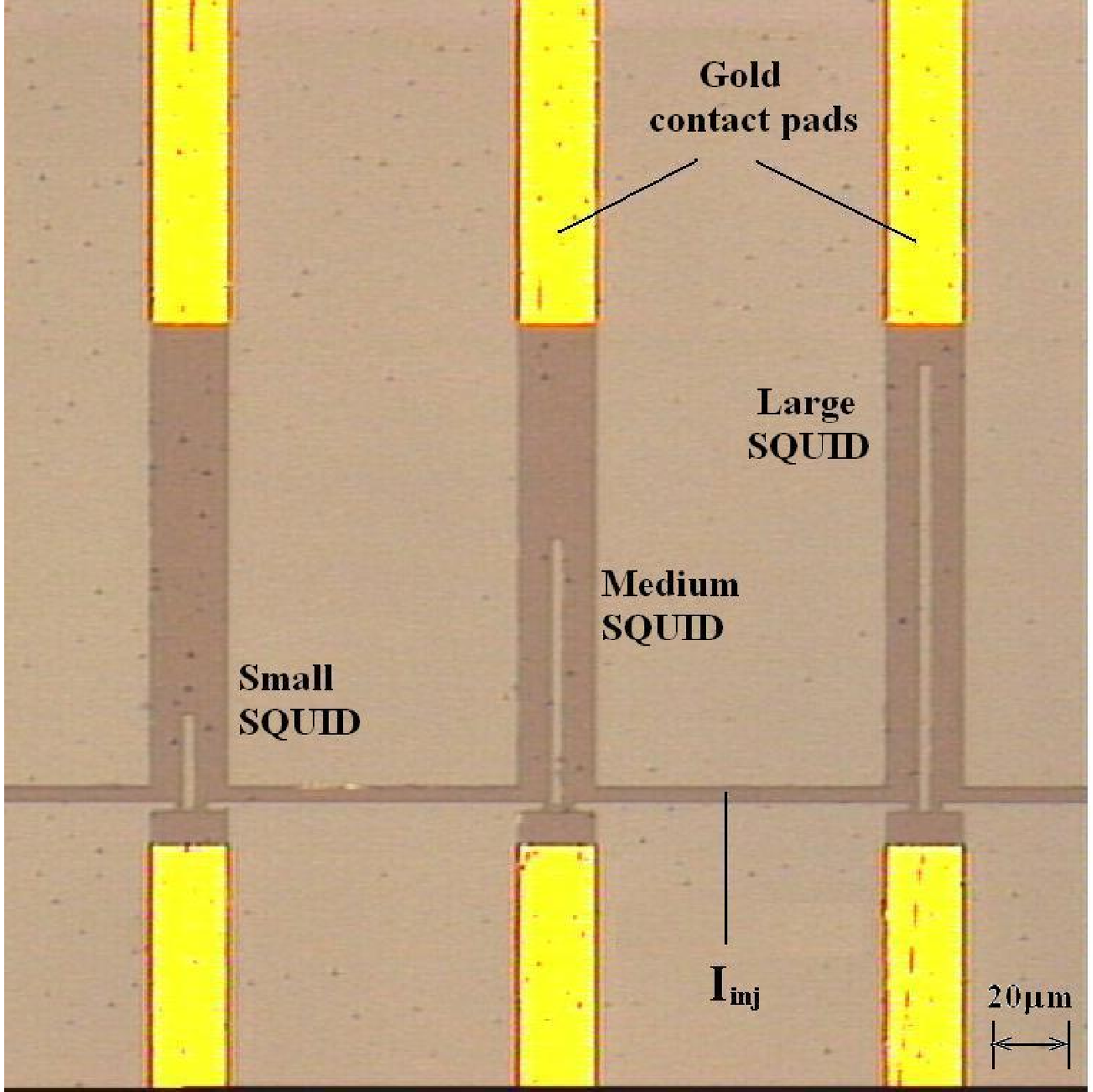}\\
  \caption{Color online. Photography of three asymmetric high-Tc DC SQUIDs.
 }\label{photo}
\end{figure}

The principal layout of an asymmetric high-T$_C$ DC SQUID is shown
in Fig. \ref{design}. It consists of the interferometer loop and
two Josephson junctions with different widths. A total of 20 such
SQUIDs was integrated on one chip. All have the same
superconducting loop width of 8 $\mu$m and slit width of 4 $\mu$m,
respectively.
Three different slit lengths of 25 $\mu$m, 75 $\mu$m and 125
$\mu$m were used, what partitions the SQUIDs into three groups:
small (S SQUID), medium (M SQUID) and large ones (L SQUID). The
slit lengths are reflected in SQUID inductance ranges of
L$\sim$65-75 pH, L$\sim$125-135 pH and L$\sim$185-195 pH for the
S, M and L SQUIDs, respectively. Within the groups, parameters of
asymmetry were varied, using various junction widths $w_1$ and
$w_2$ between 0.4 and 1.2 $\mu$m. All SQUIDs were connected via a
common superconducting line, which was used to inject a DC current
I$_{inj}$ into the SQUIDs. Figure \ref{photo} shows a photograph
of three SQUIDs connected that way.

The SQUIDs were fabricated using a standard thin-film technology,
which was described in detail in Ref. \onlinecite{IJss1}. Using
laser ablation, a 140 nm thick YBCO film is deposited onto a
24$^0$ MgO bicrystal substrate with 10 mm x 10 mm x 1 mm
dimension. On top of the non-structured YBCO film a 100 nm thick
gold layer was deposited by thermal evaporation and subsequently
structured by lift-off. Finally, the YBCO layer was patterned
using standard e-beam lithography and ion-beam-etching. As already
shown in \cite{IJss1}, Josephson junctions with widths down to 0.4
$\mu$m show no degradation in critical current density. In
addition, the chip was covered by a Teflon layer which protected
it from water during thermo cycling \cite{IJss2}. Then the SQUID
chip was glued on a non-metallic holder with a wire-wound coil
beneath, which was used to apply magnetic flux to the SQUIDs.

All measurements were performed in liquid nitrogen at 77 K in a
well shielded environment. First, for every SQUID we have measured
its voltage-current (VCC) and voltage-flux (VFC) characteristics.
Second, one of the two junctions in the loop was measured
directly. For that after all the measurements at the SQUIDs have
been performed, all SQUID loops have been cut to allow the
independent measurements of the characteristics of one junction
per SQUID. To do that the right superconducting arm of every SQUID
loop was removed by additional optical lithography and chemical
etching. This allowed to measure the critical current and normal
resistance of the Josephson junction in the remaining arm of the
SQUID. Therefore, we determined for the first time independently
and directly the current density and resistivity for every
junction in the interferometer loop which allowed us to
investigate their influence on the SQUID characteristics.

\subsubsection{SQUID characteristics}
The first group of measurements concerned the characterization and
classification of the SQUIDs. All these parameters are listed in
Table \ref{tab:table1}.
\begin{figure}
  \includegraphics[width=7 cm]{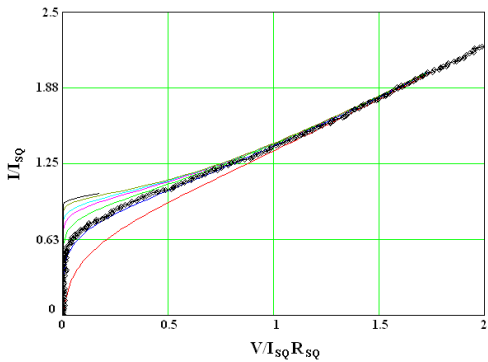}\\
  \caption{Color online. Determination of critical current from experimental VCC (thick
  line). The critical current $I_C$ to which the experimental VCC
  is normalized, is varied until the normalized experimental VCC fits
  to one of the theoretical VCCs (thin lines).
  }\label{VCC}
\end{figure}

\begin{table}
\centering \caption{\label{tab:table1}  DC SQUID parameters;
$w_1$, $w_2$ are junctions' widths; $\alpha_g$ accounts for
geometric asymmetry; L, I$_{SQ}$, R$_{SQ}$ are the DC SQUID
inductance, its critical current and normal resistance; $\Delta
V_{max}$ is the maximum swing of the VFC; I$_{max}$ is the bias
current which provides $\Delta V_{max}$. }
\begin{tabular}{c|c|c|c|c|c|c|c}
\hline\\
SQUID\#group & $w_1$,$w_2$ & $\alpha_g$ & \emph{L} & $I_{SQ}$ & R$_{SQ}$ & $\Delta V_{max}$ & $I_{max}$ \\
      &  $\mu$m && pH  & $\mu$A & $\Omega$ & $\mu$V & $\mu$A \\
\hline \hline
1 S & 0.6, 0.6 & 0 & 71.4 & 36 & 3.95 & 35.8 & 29.5\\

\hline
2 S & 0.6, 0.6 & 0 & 71.4 & 22 & 5.79 & 48.2 & 22.1 \\

\hline
3 S & 0.8, 0.4 & 0.33 & 73.8 & 35 & 2.99 & 21.8 & 35.3 \\

\hline
4 S & 0.8, 0.8 & 0 & 65.4 & 29.5 & 4 & 38.8 & 25.6 \\

\hline
5 S & 1, 0.6 & 0.25 & 66.6 & 25.5 & 4.82 & 43.8 & 23.1 \\

\hline
6 S & 1.2, 0.4  & 0.5 & 70.6 & 42 & 3.4 & 35.6 & 36.1 \\

\hline\hline
7 M & 0.6, 0.6 & 0 & 131.6 & 31 & 4.93 & 21.5 & 28 \\

\hline
8 M & 0.6, 0.6 & 0 & 131.6 & 33 & 3.87 & 16.4 & 26.3 \\

\hline
9 M & 0.8, 0.4  & 0.33 & 134.1 & 40 & 4.56 & 16.4 & 32.1 \\

\hline
10 M & 0.8, 0.8 & 0 & 125.7 & 30.5 & 4.11 & 16.9 & 25 \\

\hline
11 M & 0.8, 0.8  & 0 & 125.7 & 56.5 & 2.61 & 12.2 & 51.2 \\

\hline
12 M & 1, 0.6  & 0.25 & 126.8 & 55 & 3.1 & 15.1 & 45.1 \\

\hline
13 M & 1, 0.6  & 0.25 & 126.8 & 35 & 3.27 & 14.5 & 29.7  \\

\hline
14 M & 1.2, 0.4  & 0.5 & 130.5 & 42 & 3.55 & 15.5 & 39.5 \\

\hline
15 M & 1.2, 0.4  & 0.5 & 130.5 & 53 & 2.76 & 13.7 & 48.3 \\

\hline\hline
16 L & 0.6, 0.6  & 0 & 191.5 & 29 & 4.34 & 9.2 & 24.2 \\

\hline
17 L & 0.8, 0.4  & 0.33 & 194.1 & 25.5 & 4.68 & 8 & 19 \\
\hline
18 L & 0.8, 0.8 & 0 & 185.8 & 46 & 3.41 & 6.7 & 30.4 \\

\hline
19 L & 1, 0.6  & 0.25 & 186.7 & 35 & 3.56 & 8.4 & 31.5 \\

\hline
20 L & 1.2, 0.4  & 0.5 & 190.9 & 49 & 2.87 & 4.9 & 41.1 \\

\hline
\end{tabular}
\end{table}
Within the groups (S, M, L) the junction sizes $w_1$ and $w_2$ are
varied between 0.4 and 1.2 $\mu$m, yielding the geometrical
asymmetry $\alpha_g$. Here, the junction widths are layout
parameters. The resulting SQUID inductance L was actually
determined, however. This was done in the following way.

The injection current I$_{inj}$ flows around the upper part of the
SQUID loop (see Fig. \ref{design}). This coupling part of the
SQUID has an inductance of L$_c$ which can be measured, because
the injection current produces a magnetic flux $\Phi_c$=L$_c$
I$_{inj}$ in the SQUID loop. The measured voltage-flux
characteristics is periodical in $\Phi_0$, what gives the needed
assignment between measured flux and injected current. Thus, the
coupling inductance can be determined. In the next step, the
coupling inductance was calculated with an algorithm described
in\cite{Hild}, which takes into account geometrical and kinetic
inductance as well. The London penetration depth $\lambda_L$ of
the high-T$_c$ film acts as a fitting parameter to meet the
measured coupling inductance. This $\lambda_L$ could be determined
very consistently to (462 $\pm$ 11) nm for all the 20 SQUIDs. Now,
using this known London penetration depth, the whole SQUID
inductance could be reliably calculated in the same way.

The normal resistance of the SQUIDs, R$_{SQ}$ was determined from
the steepness of the resistive part of VCC at large currents (I
$>$ 10 I$_{SQ}$). Now, knowing the normal resistance of the SQUID,
its critical current I$_{SQ}$ can be assessed. The direct
extraction from the experimental VCC is difficult due to the large
level of thermal noise, which gives strong rounding of the curves.
To get access to the noise-free intrinsic critical current of the
SQUID, a set of theoretical VCCs is calculated, which were
obtained from an analytical solution of the SQUID equations in the
small inductance limit\cite{Ambeg} (see Fig. \ref{VCC}). In
comparison to this set of theoretical VCCs (current normalized to
I$_{SQ}$ vs. voltage normalized to the I$_{SQ}$R$_{SQ}$ product)
the experimental VCC is plotted. Because R$_{SQ}$ is known, the
noise-free critical current I$_{SQ}$ can be tuned until the
experimental curve fits well to any of the theoretical ones.

\begin{figure}
  \includegraphics[width=7 cm, angle=-90]{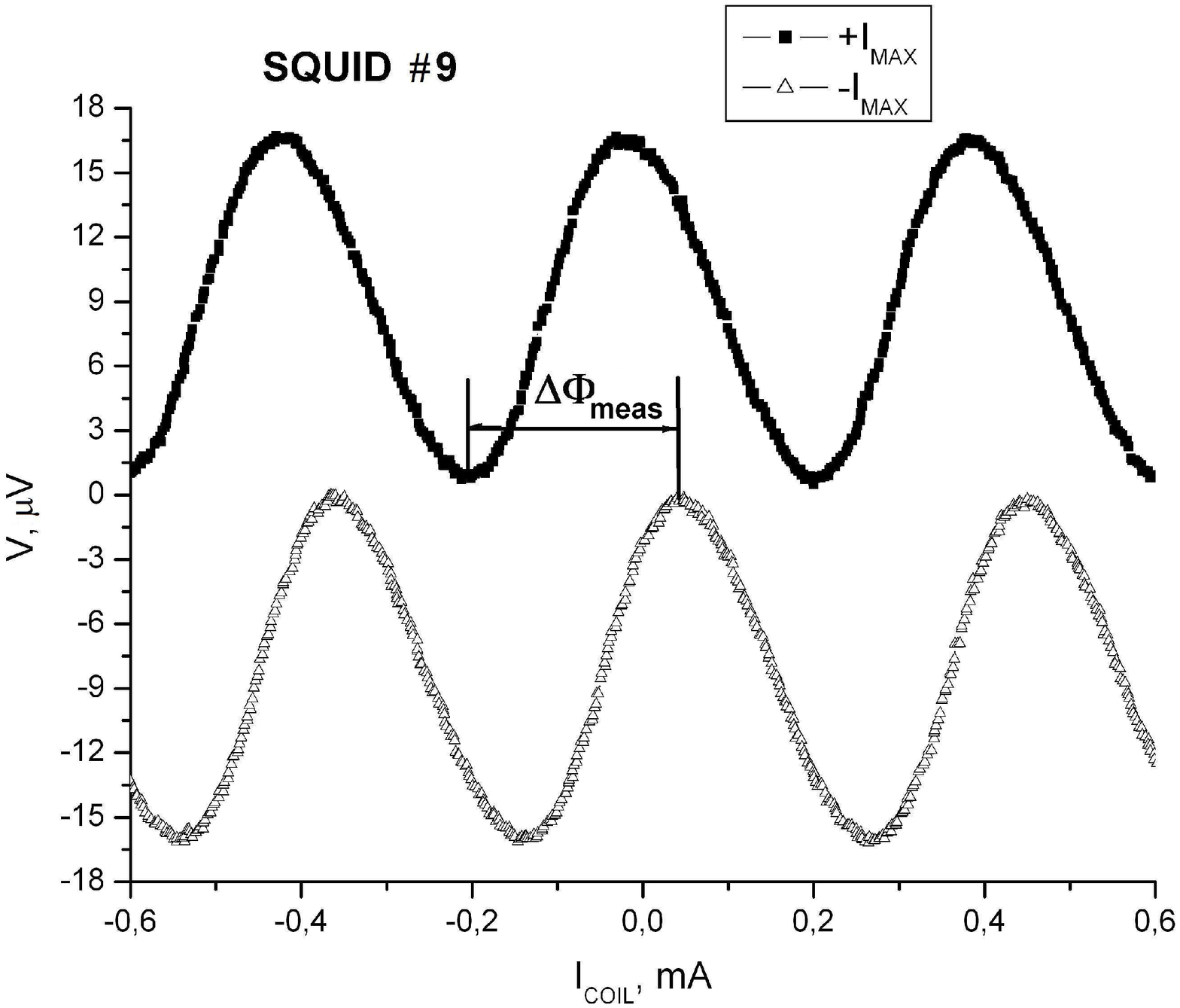}\\
  \caption{Experimental VFC for SQUID \#9. A measured flux shift $\Delta\Phi_{meas}$
   is determined as a distance between the first extremums of VFCs relative to
   zero flux point ($I_{coil}=0$).}\label{fig7}
\end{figure}

\begin{table}
\centering \caption{\label{tab:table2}  DC SQUID parameters for
the bias current $I_{max}$ given in Table I, where maximum swing
$\Delta V_{max}$ is provided; $\Delta\Phi_{max}$ is the flux shift
of the VFC under reversal of $I_{max}$; $V_{\Phi}^{\pm}$
represents the maximum steepness of the VFC at their left (+) and
right (-) sides; $\Delta V_{max}/R$ is the maximum voltage swing
related to the average junction resistance (see Table III). }

\begin{tabular}{cccccc}
\hline\\
SQUID\# group&  $\Delta\Phi_{max}$ & $V_{\Phi}^+$ & $V_{\Phi}^-$ & $\Delta V_{max}/R$ \\
      &   $\Phi_0$ & $\mu V/\Phi_0$ & $\mu V/\Phi_0$ & $\mu V/\Omega$ \\
\hline\hline
1 S&   0.357 & 115.7 & -99.7 & 4.531 \\

\hline
2 S&   0.532 & 170.1 & -147.3 & 4.163  \\

\hline
3 S&   0.675 & 71.3 & -60.01 & 5.234  \\
\hline
4 S&   0.23 & 129.5 & -116 & 4.855 \\

\hline
5 S&   0.412 & 120.13 & -167.3 & 4.548 \\

\hline
6 S&  1.243 & 102.89 & -121.4 & 5.236   \\

\hline\hline
7 M&  1.36 & 74.64 & -59.97 & 2.183  \\

\hline
8 M&   0.079 & 50.74 & -48.11 &2.123 \\

\hline
9 M&   1.656 & 57.49 & -41.19 & 1.798 \\

\hline
10 M&   -1.124 & 52.44 & -51.26 & 2.056  \\

\hline
11 M&   0.81 & 36.57 & -37.72 & 2.341  \\

\hline
12 M&   0.842 & 45.53 & -45.08 & 2.437 \\

\hline
13 M&   1.344 & 49.85 & -37.34 & 2.215 \\

\hline
14 M&   1.985 & 52.38 & -41.92 & 2.187 \\

\hline
15 M&   1.792 & 41.95 & -34.51 & 2.482  \\

\hline\hline
16 L&   0.947 & 22.6 & -22.29 & 1.060  \\

\hline
17 L&   0.948 & 22.58 & -25.24 & 0.851 \\

\hline

18 L&  -0.415 & 17.74 & -20.19 & 0.979   \\

\hline
19 L&   1.455 & 23.96 & -23.69 & 1.181  \\

\hline
20 L&   2.934 & 13.49 & -15.79 & 0.850 \\

\hline
\end{tabular}
\end{table}
The maximum voltage modulation $\Delta V_{max}$ and the bias
current I$_{max}$ needed for that voltage swing could directly be
taken from the measurement of the voltage-flux characteristics.

At the bias current I$_{max}$, which gave the maximum swing of the
VFC (both shown in Table \ref{tab:table1}), the next SQUID
parameters were measured. They are presented in Table
\ref{tab:table2}. The transfer functions $V_{\Phi}^{\pm}$
represents the maximum steepness of VFC at their left (+) and
right (-) sides. As VFC is a $\Phi_0$ periodic function, a unique
determination of the actual shift $\Delta\Phi_{max}$, which is
given in the second column of Table \ref{tab:table2}, is not, in
general, possible. Below in Section IVB we describe in detail a
reasonable procedure which relates $\Delta\Phi_{max}$ with
$\Delta\Phi_{meas}$ defined as the distance between the first
extremums of VFCs relative to zero flux ($I_{coil}=0$, see Fig.
\ref{fig7}).

The asymmetry parameters of the SQUIDs are given in Table
\ref{tab:table3}.

Unlike the method of Ref. \onlinecite{Mueller}, where the
asymmetry of the critical current was assessed indirectly from the
flux shift of the VFC, we determined the asymmetry parameters by a
direct measurement of critical current and resistance of one of
the two junctions in the SQUIDs. This got possible due to the
opening of one superconducting arm of every SQUID loop.

The first parameters given in Table \ref{tab:table3}, the critical
current I$_C$ and resistance R are average junction values for the
SQUID as defined in (\ref{curr}), (\ref{resist}). They differ from
the SQUID values given in Table \ref{tab:table1} by a factor of 2.
For a symmetric SQUID these are just the values of each of the two
junctions. The quantities I$_2$ and R$_2$ are critical current and
resistance of the second junction which remained for the
measurement after the SQUID loop had been cut. Using all these
data, we obtain the bulk asymmetries $\gamma$ and $\rho$ using
equations (\ref{curr}) and (\ref{resist}), and, finally, with the
aid of equations (\ref{intr asymm}), the intrinsic asymmetries
$\alpha_j$ and $\alpha_{\rho}$. So, for the first time the
asymmetry parameters of the current density and resistivity of the
junctions in high T$_C$ DC SQUIDs were determined independently
and directly.

\begin{table}
\centering \caption{\label{tab:table3}  Current and resistance
asymmetry of DC SQUID.}

\begin{tabular}{|c|c|c|c|c|c|c|c|c|}
\hline\\
SQUID\#  & $I_C$ & $R$ & $I_{C2}$ & $R_2$ & $\gamma$ & $\rho$ & $\alpha_j$ & $\alpha_{\rho}$\\
 group &  $\mu$A & $\Omega$  & $\mu$A & $\Omega$ & \\
\hline
1 S& 18 & 7.9 & 11 & 9.45 & 0.39 & 0.16 & -0.39 & -0.16\\

\hline
2 S& 11 & 11.58 & 9 & 13.56 & 0.18 & 0.15 & -0.18 & -0.15 \\

\hline
3 S& 17.5 & 5.98 & 7.8 &  12.09& 0.55 & 0.51 & -0.269 & -0.216 \\
\hline
4 S& 14.75 & 7.99 & 12.5 & 8.69 & 0.15 & 0.08 & -0.15& -0.08 \\

\hline
5 S& 12.75 & 9.63 & 8.5 & 13.05 & 0.33 & 0.26 & -0.087 & -0.011 \\

\hline
6 S& 21 & 6.8 & 12 & 12.77 & 0.43 & 0.47 & 0.089 & 0.039 \\

\hline\hline
7 M& 15.5 & 9.85 & 4.2 & 23.84 & 0.73 & 0.59 & -0.73 & -0.59\\

\hline
8 M& 16.5 & 7.73 & 15 & 8.32 & 0.09 & 0.07 & -0.09 & -0.07\\

\hline
9 M& 20 & 9.12 & 3 & 18.77 & 0.85 & 0.51 & -0.723 & -0.216\\

\hline

10 M& 15.25 & 8.22 & 23 & 5.26 & -0.51 & -0.56 & 0.51 & 0.56 \\

\hline
11 M& 28.25 & 5.21 & 20.5 & 6.03 & 0.28 & 0.14 & -0.28 & -0.14 \\

\hline
12 M& 27.5 & 6.2 & 24 & 7.13 & 0.13 & 0.13 & 0.124 & 0.124 \\

\hline
13 M& 17.5 & 6.54 & 5.9 & 13.71 & 0.76 & 0.52 & -0.63 & -0.31 \\

\hline
14 M& 21 & 7.09 & 3.6 & 25.74 & 0.83 & 0.72 & -0.564 & -0.344 \\

\hline
15 M& 26.5 & 5.52 & 9.8 & 12.43 & 0.63 & 0.56 & -0.19 & -0.083 \\
\hline\hline

16 L& 14.5 & 8.68 & 6.5 & 13.21 & 0.55 & 0.34 & -0.55 & -0.34\\

\hline
17 L& 12.75 & 9.35 & 5.3 & 13.62 & 0.58 & 0.31 & -0.309 & 0.022\\
\hline
18 L& 23 & 6.82 & 28.5 & 5.58 & -0.24 & -0.22 & 0.24 & 0.22 \\

\hline
19 L& 17.5 & 7.11 & 8.8 & 11.46 & 0.5 & 0.38 & -0.286 & -0.144 \\

\hline
20 L& 24.5 & 5.74 & 6.4 & 16.69 & 0.74 & 0.66 & -0.381 & -0.239 \\

\hline
\end{tabular}
\end{table}
Before we discuss the influence of junction asymmetry  on the
output SQUID characteristics we should like to show to what extent
our high $T_C$ Josephson junctions obey the scaling law
$I_CR\approx j_c^{1/2}$\cite{Gross}. In this case the parameters
$\alpha_j$ and $\alpha_{\rho}$ are interrelated to each other by
Eqs. \ref{intr asymm rel}\cite{Mueller}.
\begin{figure}
  \includegraphics[width=7 cm, angle=-90]{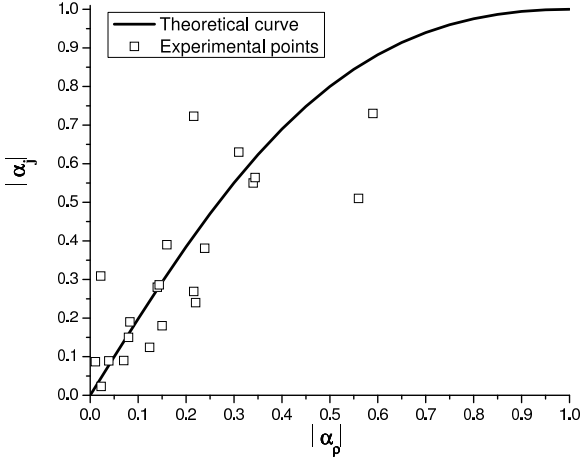}\\
  \caption{The dependance of $\alpha_j$ on $\alpha_{\rho}$. The theoretical dependance
  was calculated from the second of Eqs. \ref{intr asymm rel}.}\label{fig13}
\end{figure}
A theoretical dependance of $\alpha_j$ on $\alpha_{\rho}$ (second
of Eqs. \ref{intr asymm rel}) for this scaling law together with
experimental points is shown in Fig. \ref{fig13}. As is seen from
this figure most of the experimental points are grouped over the
theoretical curve. Therefore, we may conclude that our high $T_C$
Josephson junctions are described by the aforementioned scaling
law with a good accuracy.
\section{Discussion}
\subsection{The depth of the voltage modulation}
\begin{figure}[tbp]
  \includegraphics[width=7 cm, angle=-90]{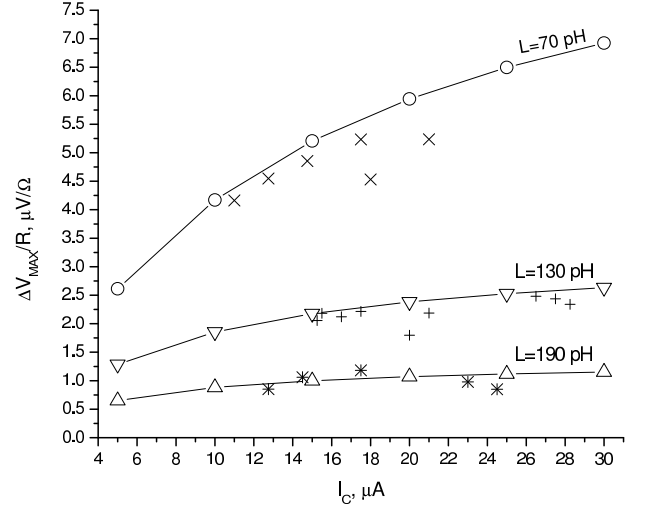}\\
  \caption{The voltage modulation as a function of the critical
  current. The lines for $L$=70 pH ($\bigcirc$), 130 pH ($\bigtriangledown$), and
  190 pH ($\bigtriangleup$) are calculated from Eq. \ref{Enp}.
  The experimental points are shown by $\times$ for
  S-SQUIDs, + for M-SQUIDs, and $\divideontimes$ for L-SQUIDs.}\label{depth_mod}
\end{figure}

In general, the increase of the SQUID inductance leads to a
decrease of the voltage swing of VFC\cite{Koelle}. This is
illustrated in Fig. \ref{depth_mod} where the voltage modulation
$\Delta V_{max}/R$ is shown as a function of the junction critical
current. As is seen from the figure the SQUIDs are grouped by
their inductance near corresponding lines obtained from the
expression of Enpuku\cite{Enp}.
\begin{equation}\label{Enp}
\frac{{\Delta V }}{{I_C R}} = \frac{4}{{\pi (1 + \beta )}}\exp
\left( { - 3.5\pi ^2 \frac{{k_B TL}}{{\Phi _0^2 }}} \right)
\end{equation}

\begin{figure}[tbp]
  \includegraphics[width=7 cm, angle=-90]{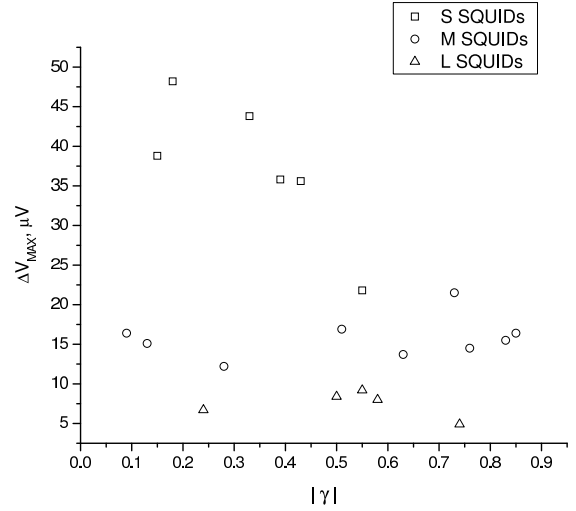}\\
  \caption{The depth of the voltage modulation as a function of the critical
  current asymmetry $\gamma$.}\label{fig9}
\end{figure}

\begin{figure}[tbp]
  \includegraphics[width=7 cm, angle=-90]{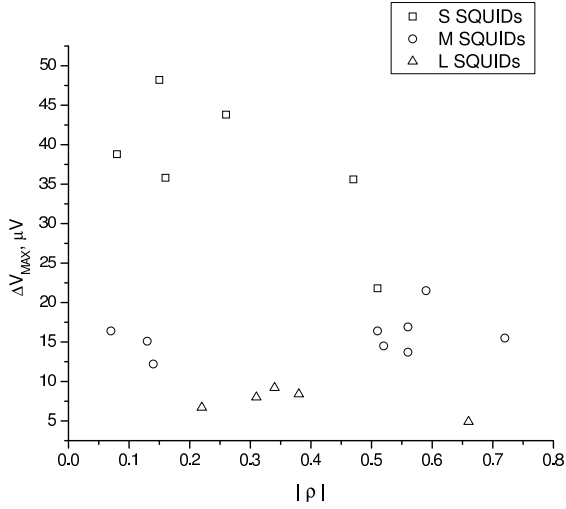}\\
  \caption{The depth of the voltage modulation as a function of the
  resistance asymmetry $\rho$.}\label{fig10}
\end{figure}
We also investigated the dependance of the maximum voltage
modulation on the asymmetry parameters of the DC SQUID (see Figs.
\ref{fig9} and \ref{fig10}). We have found that for M- and
L-SQUIDs the influence of junction asymmetry on the voltage
modulation is weaker than that for small inductance SQUIDs. For S
SQUIDs  there is appreciable scattering of $\Delta V_{max}$ versus
$\gamma$ and $\rho$.
\subsection{Flux shift}
\begin{figure}[h]
  \includegraphics[width=7 cm, angle=-90]{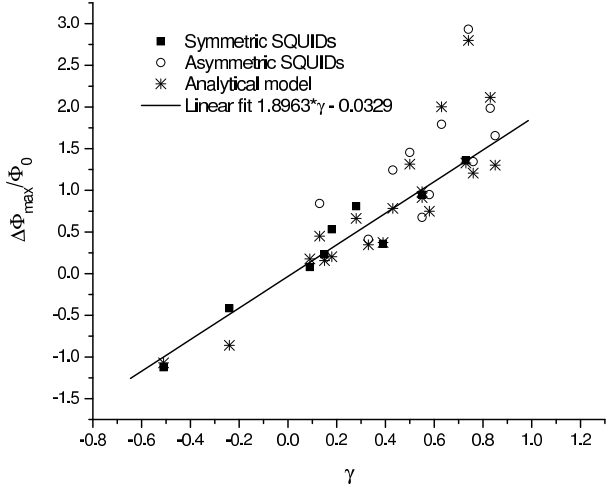}\\
  \caption{The flux shift as a function of the
 current asymmetry $\gamma$. Geometrically symmetric SQUIDs ($\alpha_g=0$) are
 shown by black boxes, SQUIDs with $\alpha_g\neq 0$ are shown by open circles.
 Theoretical points calculated from (\ref{shift3}) are shown by stars.}\label{fig11}
\end{figure}

  \begin{figure}[h]
  \includegraphics[width=7 cm, angle=-90]{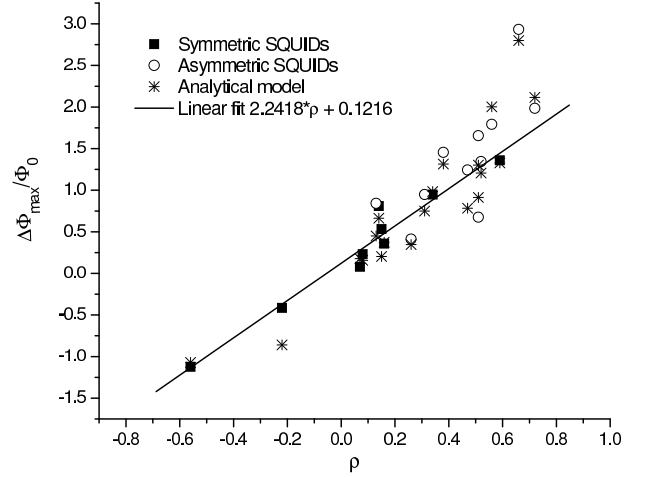}\\
  \caption{The flux shift as a function of the
  resistance asymmetry $\rho$. Geometrically symmetric SQUIDs ($\alpha_g=0$) are
 shown by black boxes, SQUIDs with $\alpha_g\neq 0$ are shown by open circles.
 Theoretical points calculated from (\ref{shift3}) are shown by stars.}\label{fig12}
\end{figure}
As is known the asymmetry of the SQUID junctions results in a flux
shift of the VFC under bias reversal. Here we determine the flux
shift as the sum of the shifts of two VFCs, one for $I=+I_{max}$
and the other for $I=-I_{max}$, relative to the zero flux point
($I_{coil}=0$ in Fig. \ref{fig7}). Here the subscript $"max"$
means the bias current which provides maximum swing of the VFC.
Since the VFC is $\Phi_0$ periodic, the flux shift can be
determined with an accuracy $2n\Phi_0$, where $n$ is an integer.
Thus the minimum uncertainty of the determination of the flux
shift for asymmetrical SQUIDs is $2\Phi_0$.

In an ideal case when there is no parasitic trapped flux in the
loop, both VFCs are shifted symmetrically by the same amount
$\Delta\Phi_{sym}$ in opposite directions with respect to zero
flux point. In this case the overall flux shift is calculated as a
sum of both shifts, $\Delta\Phi=2\Delta\Phi_{sym}$. However, in
our cases the shift of two VFCs is not symmetric with respect to
zero point (see Fig. \ref{fig7}). The reason for this is the
parasitic flux $\Phi_p$ trapped in a loop. This flux does not
change its sign under bias reversal, hence the flux shifts are as
follows: $\Delta\Phi_-=\Delta\Phi_{sym}+\Phi_p$,
$\Delta\Phi_+=-\Delta\Phi_{sym}+\Phi_p$, where $\Delta\Phi_-$ is
the flux shift for the VFC for reversed bias current
($I=-I_{max}$), and $\Delta\Phi_+$ is the same quantity for the
unreversed VFC ($I=+I_{max}$). Thus, the parasitic flux disappears
from the overall flux shift
$\Delta\Phi=\Delta\Phi_--\Delta\Phi_+=2\Delta\Phi_{sym}$. The
application of this simple formula requires precise knowledge of
the corresponding extremums of the VFCs shifted by a parasitic
flux. In other words, we must know these two extremum points on
both VFCs whose distance provides us with the overall flux shift.
However, since VFC is $\Phi_0$ periodic, it is not possible to
definitely determine these two points. Therefore, a single valued
determination of the flux shift requires some additional
information.

The procedure we used for the unique determination of the actual
flux shift, $\Delta\Phi_{act}$ was performed in three steps.
First, we measured the flux shift $\Delta\Phi_{meas}$ between the
\emph{nearest} extremum points of VFCs with respect to the zero
flux point. Second, by using expression (\ref{shift3}) we
calculated for a given SQUID a theoretical value of the flux shift
$\Delta\Phi_{theor}$. Finally, the actual flux shift
$\Delta\Phi_{act}$ was determined by subtracting from (or adding
to) $\Delta\Phi_{meas}$ the integer number of flux quanta
$n\Phi_0$ ($n=0,\pm 1,\pm 2, etc.$) which gave the flux shift
nearest to the theoretical value $\Delta\Phi_{theor}$.

The illustration of our method is presented in Table \ref{flux
shift determination}.  The values of actual flux shift from the
last column of this table are given as $\Delta\Phi_{max}$ in the
second column of Table \ref{tab:table2}.

We want to stress here that in our method the theoretical
estimations of the flux shift were used only as a guide for the
determination of the actual flux shift caused by the predetermined
asymmetry. This differs from many other papers on the subject
(see, for example, Ref. \onlinecite{Mueller}) where theoretical
expression (\ref{shift1}) itself has been used for the
determination of asymmetry parameters. In this latter case there
are not any means which allow a unique discrimination between the
unshifted curve and the curve shifted by a parasitic flux.

\begin{table}
  \centering
  \caption{Determination of actual flux shift. $\Delta\Phi_{meas}$ is the
flux shift measured between nearest extremums of two VFCs,
$\Delta\Phi_{theor}$ is the flux shift calculated from Eq.
\ref{shift3}, and $\Delta\Phi_{act}$ is the actual flux shift
corrected for parasitic flux.}\label{flux shift determination}
\begin{tabular}{|c|c|c|c|}
\hline\\
SQUID\# group & $\Delta\Phi_{meas}$ & $\Delta\Phi_{theor}$ &$\Delta\Phi_{act}$  \\
     &  $\Phi_0$ &$\Phi_0$, Eq.(7)&  $\Phi_0$\\
 \hline \hline
1 S & 0.357 & 0.374 & 0.357 \\

\hline
2 S & 1.532 & 0.203 & 1.532-1=0.532 \\

\hline
3 S & 1.675 & 0.913 &1.675-1=0.675 \\

\hline
4 S & 1.23 & 0.157 & 1.23-1=0.23  \\

\hline
5 S & 0.412 & 0.347 & 0.412 \\

\hline
6 S &1.243  & 0.784 & 1.243  \\

\hline\hline
7 M & 1.36 & 1.328 & 1.36  \\

\hline
8 M & 0.079 & 0.177 & 0.079  \\

\hline
9 M & 0.656  & 1.302 & 0.656+1=1.656  \\

\hline
10 M & -1.124 & -1.07 & -1.124  \\

\hline
11 M & 0.81  & 0.663 & 0.81  \\

\hline
12 M & 0.842 & 0.451 & 0.842  \\

\hline
13 M & 1.344  & 1.207 & 1.344   \\

\hline
14 M & 0.985  & 2.116 & 0.985+1=1.985  \\

\hline
15 M & 0.792  & 2.002 & 0.792+1=1.792  \\

\hline\hline
16 L & 0.947  & 0.984 & 0.947 \\

\hline
17 L & 0.948  & 0.75 & 0.948  \\
\hline
18 L & -0.415 & -0.86 & -0.415  \\

\hline
19 L & 0.455  & 1.315 & 0.455+1=1.455 \\

\hline
20 L & 0.934  & 2.8 & 0.934+2=2.934 \\

\hline
\end{tabular}
\end{table}

The dependance  of experimental flux shift (second column in Table
\ref{tab:table2}) on current asymmetry $\gamma$ and resistance
asymmetry $\rho$ is given in Figs. \ref{fig11} and \ref{fig12}.
Also we show on these graphs  the theoretical points (star
symbols) which were calculated from Eq. (\ref{shift3}) by using
the measured values of $\gamma$ and $\rho$. As is seen from these
graphs, the equation (\ref{shift3}) is a good approximation for
the measured flux shift. The experimental points for the flux
shift can be approximated by least mean square fits which are as
follows: $\Delta\Phi_{max}/\Phi_0=1.89\gamma-0.03$ for Fig.
\ref{fig11} and $\Delta\Phi_{max}/\Phi_0=2.24\rho+0.12$ for Fig.
\ref{fig12}.

\subsection{Transfer function}
\begin{figure}
  \includegraphics[width=7 cm, angle=-90]{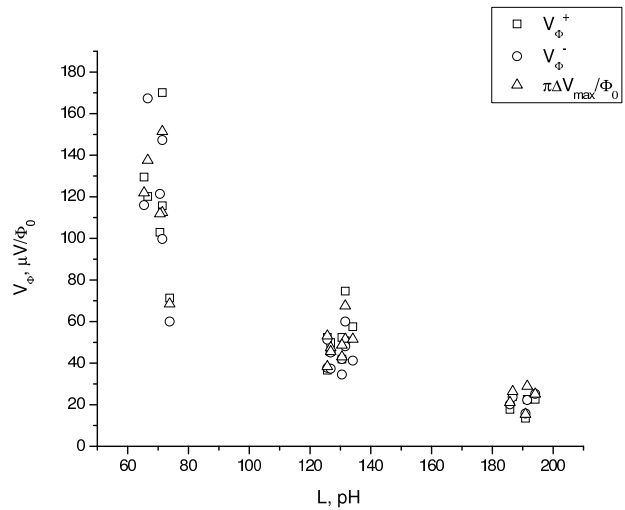}\\
  \caption{Dependance of the transfer functions $V_{\Phi}^{\pm}$ on SQUID inductance.}
  \label{trf squid n}
\end{figure}

\begin{figure}
  \includegraphics[width=7 cm, angle=-90]{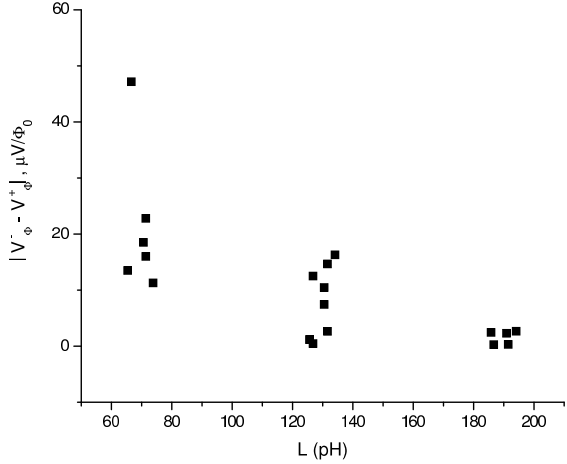}\\
  \caption{The deviation of VFC shape from symmetrical form.}\label{deviation}
\end{figure}

For every SQUID the transfer functions $V_{\Phi}^{\pm}$ were
measured (see Table \ref{tab:table2}) and their dependance on
different SQUID parameters was investigated. The dependance of
$V_{\Phi}^{\pm}$ on the loop inductance $L$ is shown in Fig.
\ref{trf squid n}. We also show on this plot the transfer function
for a symmetric shape of the VFC by using the relation
$V_{\Phi}=\pi\Delta V/\Phi_0$ with $\Delta V$ being taken from
Table \ref{tab:table1}. It is seen that, in general, the increase
of inductance results in a decrease of both  the transfer
functions and the scattering for the transfer function values. The
largest scattering is observed for S SQUIDs. Another conclusion is
that as the SQUID inductance is increased the shape of VFC becomes
more symmetric: the transfer function for symmetric shape becomes
closer to the measured transfer functions $V_{\Phi}^{\pm}$. As a
measure of the shape asymmetry of VFC we introduce the quantity
$|V_{\Phi}^+-V_{\Phi}^-|$ which is equal to zero for a symmetrical
shape. The dependance of this quantity on SQUID inductance is
shown in Fig. \ref{deviation}. From this figure it is clearly seen
that the shape of VFC becomes more symmetric as the SQUID
inductance becomes larger.

\begin{figure}
  \includegraphics[width=7 cm, angle=-90]{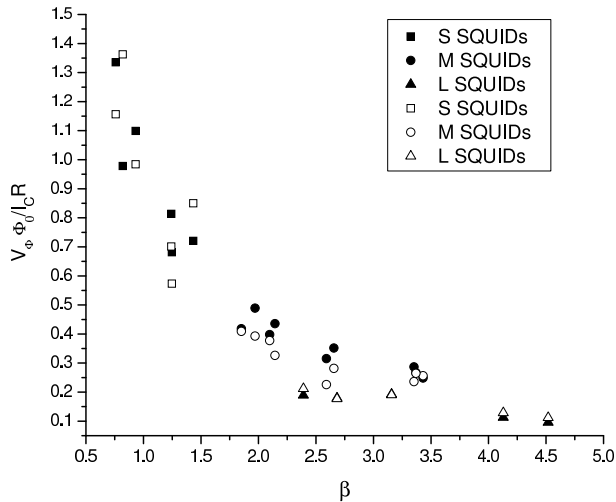}\\
  \caption{Dependance of the transfer functions
on $\beta$. Black symbols are for $V_{\Phi}^{+}$, white symbols
are for $V_{\Phi}^{-}$.}\label{trf beta}
\end{figure}

The dependance of the measured transfer functions $V_{\Phi}^{\pm}$
on $\beta$ is shown in Fig. \ref{trf beta}. Here one can see a
nearly exponential decay of the transfer function with the
increase of $\beta$.

The dependance of the normalized transfer functions
$V_{\Phi}^{\pm}\Phi_0/RI_C$ on current asymmetry $\gamma$ and
resistance asymmetry $\rho$ is shown in Figs. \ref{trf gamma} and
\ref{trf rho}. As is seen from these figures, the L SQUIDs show a
weak dependance on the parameters of asymmetry. It conforms with
Fig. \ref{deviation}, where the VFCs for these SQUIDs have minimum
distortion. However, for SQUIDs with small inductance (S SQUIDs)
the dependance on asymmetry parameters is much stronger. It is
worth noting that the $\gamma$ and $\rho$ dependances of
$V_{\Phi}^{\pm}$ (see Figs. \ref{trf gamma}, \ref{trf rho}) have
indications of a peak for S SQUIDs near the value of 0.2. A
similar behavior has been found for SQUIDs with $L\approx 20$ pH
in Ref. \onlinecite{Mueller} (see Fig. 6 therein).

We also compare the dependance of measured transfer functions
$V_{\Phi}^{+}$ on current asymmetry $\mid\alpha_J\mid$ for
geometrically symmetric SQUIDs ($\alpha_g=0$) with the
corresponding expressions  from Ref. \onlinecite{Mueller} obtained
from computer simulations. The results are shown in Fig. \ref{trf
symm}. Again, the strongest deviation of experimental points from
theoretical ones is observed for S SQUIDs (only SQUID \#7 is an
exception).
\begin{figure}
  \includegraphics[width=7 cm, angle=-90]{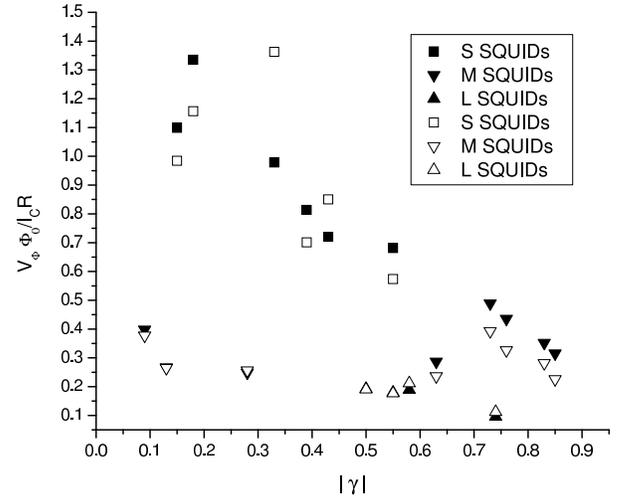}\\
  \caption{Dependance of the transfer functions
on current asymmetry $\gamma$. Black symbols are for
$V_{\Phi}^{+}$, white symbols are for $V_{\Phi}^{-}$.}\label{trf
gamma}
\end{figure}
\begin{figure}
  \includegraphics[width=7 cm, angle=-90]{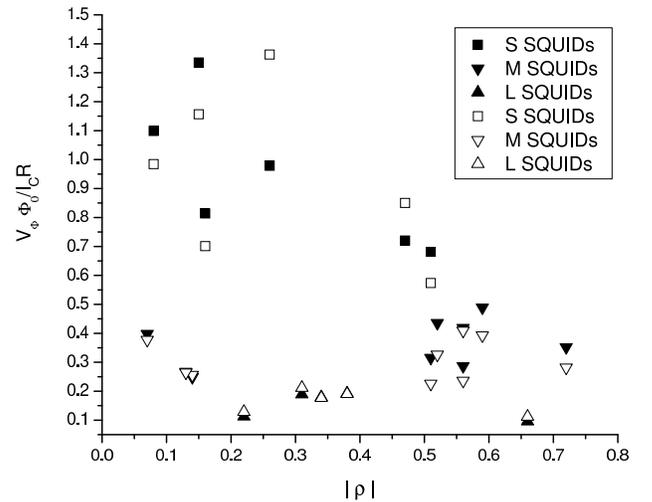}\\
  \caption{Dependance of the transfer functions
on resistance asymmetry $\rho$. Black symbols are for
$V_{\Phi}^{+}$, white symbols are for $V_{\Phi}^{-}$.}\label{trf
rho}
\end{figure}
\begin{figure}
  \includegraphics[width=7 cm, angle=-90]{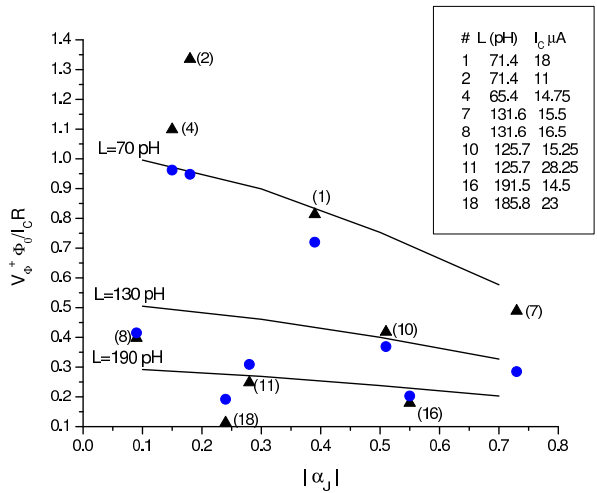}\\
  \caption{Color online. The transfer function $V_{\Phi}^{+}$ for geometrically
  symmetric SQUIDs ($\alpha_g=0$). Black triangles with SQUID \#
  are experimental points;  blue circles are corresponding values
  calculated by using Eqs. 6 for intrinsic asymmetry from Ref.\onlinecite{Mueller}.
  Solid lines show a guide dependance for three different inductances
  calculated from the same equations for $I_C=15\mu A$. The insert shows
  the values of inductance and critical current for SQUIDs with $\alpha_g=0$ taken from
  Table \ref{tab:table1}.} \label{trf symm}
\end{figure}
\section{Conclusion}
The main purpose of the study was the experimental investigation
of the influence of the junction asymmetries on the SQUID output
characteristics (depth of modulation, flux shift, transfer
function). To this end we directly measured the current and
resistance asymmetry of the junctions in every of 20 investigated
SQUIDs. It turned out that the values of current and resistance
asymmetry ($\alpha_j$ and $\alpha_{\rho}$) were randomly
distributed over a chip without noticeable correlation either with
SQUID inductance or critical current. Even for geometrically
symmetric SQUIDs (we had 9 such SQUIDs) there was a significant
asymmetry in critical current and resistance. Nevertheless we
achieved definite conclusions about the influence of the junction
asymmetry on the SQUID output characteristics.

The dependance of the depth of modulation $\Delta V$ and the
transfer functions $V_{\Phi}^{\pm}$ on the junction asymmetry is
appreciable only for low inductance SQUIDs (see Figs. \ref{fig9},
\ref{fig10}, \ref{trf gamma}, \ref{trf rho}). For large
inductances this dependance is rather weak which is in accordance
with a small distortion of the shape of VFC for these SQUIDs (see
Fig. \ref{deviation}). Therefore, we may conclude that, in
general, both the voltage modulation and the transfer function are
not very sensitive to the junctions asymmetry.  However, for
SQUIDs with a relatively small inductance $L<70 $ pH the
dependance on asymmetry is more significant, the shape of VFC is
more distorted and the transfer functions are well above the
corresponding values for large inductance SQUIDs.

As was expected, the flux shift of the VFC is more sensitive to
the junction asymmetry than the depth of modulation or transfer
functions. The dependance of the flux shift on $\gamma$ and $\rho$
is approximately linear and it is well described by the analytical
model (Figs. \ref{fig11}, \ref{fig12}).

The results obtained in the paper are important for the
implementation in the sensitive instruments based on high T$_C$
SQUID arrays and gratings.

\section*{Acknowledgment}
The authors are indebted to V. Zakosarenko for helpful discussions
related to the measurements, and to M. Sondermann and S. Giessler
for the preparation of SQUID chips for the measurements. Ya. S. G.
thanks J. T. Jeng for useful comments and for providing the
manuscript of his contribution to ASC'08 prior its publication. I.
L. N. acknowledges the hospitality of the Institute of Photonic
Technology, Jena, Germany, where the experimental part of the work
has been performed. The work was partly supported by DAAD and
Ministry of Education and Science of Russian Federation (Programm
"Michail Lomonosov", Grant \# 9686).

\end{document}